\documentclass[pra,amsmath,amssymb,showpacs]{revtex4}
\usepackage{epsfig}

\begin{document}

\title{A Tale of Two Representations: Energy and Time in Photoabsorption}


\date{\today}

\author{A. R. P. Rau\footnote{e-mail: arau@phys.lsu.edu}}

\affiliation{Department of Physics and Astronomy, Louisiana State University, Baton Rouge, Louisiana 70803, USA}

\begin{abstract}
{\bf Abstract}
This essay is based on a talk at Advances in Atomic, Molecular, and Optical Sciences 2020 (AAMOS20) in a symposium honoring Prof. S. T. Manson's decades-long contribution to photoabsorption studies. Quantum physics introduced into physics pairs of conjugate quantities bearing a specific complementary relationship, energy and time being one such pair. This gives rise to two alternative representations, a time-dependent and a time-independent one, seemingly very different but both capable of embracing the same physics. They give complementary descriptions and insight, with technical questions, theoretical and experimental, determining which may be the more convenient and practicable at any juncture. Two recent topics, Cooper minima in photoabsorption in Cl$^-$ and Ar, and angular-momentum barrier tunneling of $f$ photoelectrons from Se in WSe$_2$, provide illustrative examples, also of the role that technological developments over the past five decades played in our approach to and understanding of phenomena. 
\end{abstract}

\maketitle

\section{Introduction}
The recognition a century ago of the quantum nature of our Universe introduced into physics that concepts such as position and momentum, linear or angular, and energy and time, that had been thought of as independent were conjugate pairs in a well-defined sense. With a universal constant, Planck's $\hbar$, characterizing our Universe, already on dimensional grounds alone, such pairs could be seen as gradients with respect to one another. Indeed, had physicists first encountered quantum and not classical physics, they might have been defined as such, that there is only energy, momentum, etc., the corresponding time and space being their gradients, respectively, with a multiplicative $\hbar$. In each pair, the opposite choice is equally valid except that since we have fundamental conservation laws of energy, momentum, and angular momentum, they could more properly be regarded as the basic entities whereas time, position, and angle coordinate, respectively, are their derived gradients in energy and momentum space.

But the conjugacy of itself raises immediately another question, especially when coupled with the definition of the state of a system. Setting aside quantum field theories, non-relativistic quantum mechanics, the domain of atomic physics discussed in this essay, describes the state as a complex entity, an abstract Dirac bra or ket, or as a wave function, which cannot simultaneously be prescribed in terms of both members of the conjugate pair. Unlike in classical mechanics of an assembly of particles where the state is defined by the coordinate positions and momenta of each of the particles, that is precluded in quantum mechanics because, once again, $\hbar$ sets a limitation on such simultaneous prescription of entities whose commutator is non-zero. Only mutually commuting entities can be fixed simultaneously and used to define physical states. This points then to choosing either one but not both of a non-commuting pair. The very first step into quantum mechanics introduces students to either a wave function in coordinate or in momentum space, even for the first elementary examples of a free particle or a harmonic oscillator. 

Similarly, either energy or time may be invoked which hearkens back already to the pairing in classical physics of frequency and time as Fourier conjugates. One can deal with a function of time or, alternatively, in Fourier conjugate space as a function of frequency. It is but one immediate next step from frequency to energy with $\hbar$ again as an agent of that translation. Waves may be described either as real standing waves of sines and cosines or as travelling waves of complex exponentials, each set satisfying the wave equation and any one of them expressible as a linear superposition of members of the other set. Stationary states of fixed energy played a central role in the development of quantum physics. Bohr's description of stationary states of the hydrogen atom accounted for both atomic sizes and energies and for their stability as per their very name. The time-dependent Schr\"{o}dinger equation for them can be reduced immediately to the real time-independent Schr\"{o}dinger equation at fixed $E$. In such a state, in keeping with the uncertainty relationship, time becomes irrelevant, the electron persisting in that state over all time. Time occurs as a background in a generally ignorable pure phase factor of the wave function.

While not necessary, with either representation capable of handling all quantum physics or atomic phenomena, it became customary to view bound states in $t$-independent terms and with real quantities, while invoking $t$ explicitly and complex quantities for collision or scattering. Even in quantum mechanics, the very words of the language we use invoke time when discussing scattering. This equally so for the toy ones of scattering from a one-dimensional potential barrier or three-dimensional scattering of electrons, protons, or other elementary particles at our most energetic accelerators. We talk of the particle, or the plane wave representing it, going in from a distant collimator source and then leaving the scatterer to be captured by detectors at infinity, possibly with other fragments. This picture is one of an evolution in time. Thus, it is almost standard and adopted without saying so for the beginning student to see bound states whether in simple potential wells or complex situations in one way, as energy eigenstates, while collisions or scattering are handled another way as evolving in time. Yet, for the real Schr\"{o}dinger equation with real energies and potentials, sines and cosines suffice for both bound states and scattering. Reflection and transmission coefficients for the latter, or phase shifts in more complicated situations, all of them real and the only experimentally accessible quantities, can be obtained without invoking time. And the analysis can be kept real throughout, the imaginary element $i$ making an appearance only through boundary conditions imposed at infinity. See, for instance, section 1.5 of \cite{ref1}. 

Spectroscopy, the domain of bound states, together with transitions between them caused by coupling to the electromagnetic field, and scattering involving asymptotic states with the electron reaching infinity, are best viewed on the same footing, as states of the same atomic Hamiltonian. Its stationary states embrace both bound and continuum states. Also, phenomena involving the latter whose asymptotic states involve the free electron include elastic or inelastic scattering and photoionization, so that they are intimately related. The last item is but ``half-scattering," differing from electron-atom/ion scattering in involving only the ``outgoing" half of electron moving from small $r$ to infinity and not the ``ingoing" bit of incident electron from infinity onto the atom/ion. Resonances are also unified into the same framework. Phase shifts in the scattering domain can be connected across ionization thresholds to quantum defects \cite{ref2}, which are parameters of the bound states on the other side of that threshold \cite{ref1,ref3}. Oscillator strengths for photoabsorption to excited bound states can be naturally continued into the photoionization region above threshold, the combined distribution providing the total response of the atomic system to radiation (Fig. 2.1 of \cite{ref1}). Many such connections between the two domains become natural in this unified view of spectroscopy and scattering \cite{ref1}.

A contrast between statics and dynamics was already made in physics and engineering long before quantum physics. The former dealt with the structure of the system on its own with time mostly irrelevant whereas dynamics dealt with motion of the physical system in space as a function of time. In retrospect, the static system could also have been considered as stationary but being translated unchanged along the time axis (counterpart of the $\exp (-iEt/\hbar)$ as a backdrop for stationary states in quantum physics). With absolute motion having no meaning in physics, only relative motion and relative velocities being meaningful, and especially with Special Relativity's placing all inertial frames on an equal footing, the distinction between statics and dynamics is artificial, all physics being transformations in space-time. See, for instance, \cite{ref4}. Quantum physics only sharpens this further. Quantum field theories go to the most extreme of seeing coordinate position and time as mere markers, while fields and their derivatives become the dynamical operators. All particles and waves are seen as excitations when the operators act on the vacuum state \cite{ref5}, with the classical limit of waves only for massless bosons, all others as particles in that limit \cite{ref6}.

\section{Phase shifts and their Derivatives}

Since its beginnings about a century ago, it has been a quest of spectroscopy in atomic physics to reach higher and higher energy resolution, both in experimental observation and in theoretical calculation. States of definite $E$ are studied with increasing precision, that is, with increasingly smaller $\Delta E$, for both bound ($E < 0$) and scattering ($E > 0$) stationary states. Synchrotrons and lasers of increasing sophistication and reach in wavelength and intensity on the experimental side and analytical and numerical calculations on the theoretical side have aimed to attain increasingly smaller values of $\Delta E$. With the non-relativistic energy-momentum relation for the free electron at infinity, $E =(\hbar k)^2/2m$, the wave vector $k$ is chosen by convention for negative energies as $k =i \kappa$. In spherical polar coordinates, with angular dependence factored out in spherical harmonics $Y_{\ell m}$ for orbital angular momentum $\ell$, a pair $(f(r), g(r))$ of regular and irregular solutions of the radial equation are asymptotically proportional to $\sin (kr -\ell \pi /2)$ and $ -\cos (kr -\ell \pi /2)$, respectively (chapter 5 of \cite{ref1}). In the presence of interactions at small and intermediate $r$, the electron's asymptotic wave function is given by a superposition of them,     

\begin{equation}
f \sin \delta_{\ell} -g \cos \delta_{\ell}   \sim  sin (kr -\frac{1}{2}\ell \pi +\delta_{\ell}), 
\label{eqn1}
\end{equation}
with phase shifts $\delta_{\ell}$ that are functions of $E$. All effects of interactions are contained in them, these being the only accessible  objects at infinity through contact with experimentally measured cross sections.

To obtain these cross sections, differential in energy, angles, spin, and other parameters, or total integrated cross sections, the full wave function for electron scattering or photoionization has to be constructed (section 7.9 of \cite{ref1}) by superposing different $\ell$ of Eq.~(\ref{eqn1}) with appropriate outgoing (+) or ingoing (-) wave boundary conditions, respectively. In enforcing these boundary conditions, the asymptotic wave function in Eq.~(\ref{eqn1}) has to be cast in terms of outgoing and ingoing spherical waves, $\exp(ikr)$ and $\exp(-ikr)$, respectively (the choice of the time dependence $\exp(-iEt/\hbar)$ implicit here even in defining the terms of out/in). The ratio of their coefficients $\exp(\pm i\delta)$, namely $\exp(2i\delta)$, is the S-matrix (sections 4.3 and 4.5 of \cite{ref1}). The similar ratio of two ``Jost matrices" applies for multi-channels (section 5.3 of \cite{ref1}). The required outgoing wave boundary condition is that the ingoing piece match the corresponding one in the incident plane wave along a particular $\vec{k}$, and all the other outgoing waves present provide the scattered piece. Thus, S-1, namely,

\begin{equation}
 e^{2i\delta} - 1 \propto e^{i\delta} \sin \delta
\label{eqn2}
\end{equation}
is the scattering amplitude. Its modulus-square, together with the dimensional element provided by $k$ and the $4\pi$ of three-dimensional space, gives the scattering cross-section (section 4.6 of \cite{ref1})  

\begin{equation}
\sigma = (4\pi/k^2) \sum_{\ell} (2\ell + 1) \sin^2 \delta_{\ell}. 
\label{eqn3}
\end{equation}
Note the weight factor and sum in Eq.~(\ref{eqn3}) to reflect isotropic degeneracy and the equal contribution of all $m$ values in an $\ell$ manifold.

Bound states when $k =i \kappa$ are given by the condition necessary to have the coefficient of $\exp(-ikr)$ vanish as required for finite, normalizable wave functions, that is, $\exp(-i\delta) = 0$. Extension to a multi-channel generalization for a complex atom is straightforward. The superposition in Eq.~(\ref{eqn1}) has in place of the sine and cosine the Jost matrices $J^{\pm}$. For scattering, the relevant combination is the so-called S-matrix of structure $J^+ (J^-)^{-1}$, and its difference from the unit matrix enters the expressions for cross sections. Photoionization is obtained through $(J^+)^{-1}$ and the multi-channel rendering of bound states is $J^- =0$. All physical observables are expressed in terms of these phase shifts or Jost matrices (section 7.3 of \cite{ref1}).
 
The phase shifts are in general $E$-dependent and contain all the information about the interaction. They, and $\tan \delta$ may be of either sign, positive for repulsive and negative for attractive interactions. Their energy dependence may be smooth and gradual or rapid over certain energy intervals, reflected correspondingly in the cross sections. A rapid climb through $\pi$ over some energy interval $\Gamma$ around $E_r$, and accompanying rise and fall of cross section, denotes a ``resonance" of width $\Gamma$ at that energy $E_r$. At this point, time may be invoked, through $\hbar/\Gamma$ as a ``lifetime" of a state at that energy. The most ready physical situation and interpretation of a resonance is of a quasi-bound state at energy $E_r$ that is embedded in a continuum and thus degenerate with it. Unlike a strict stationary bound state, there is a necessary uncertainty $\Gamma$ in its energy. Correspondingly, it may be viewed in terms of the finite lifetime of that quasi-bound state. Doubly-excited states such as $2s2p \, ^1 P^o$ of the He atom or the H$^-$ negative ion provide an example, such a state degenerate with a background continuum of singly ionized $1sEp \, ^1 P^o$. Upon exciting both electrons into such a state, say by photoabsorption from the ground state (with a photon of $\approx$ 65 eV in He and $\approx$ 11 eV in H$^-$), one of the electrons can be ejected to infinity, the other simultaneously dropping down to the ground state in the phenomenon of auto-ionization. It is the Coulomb coupling between the electrons that is responsible for this process with corresponding lifetimes of $~ 10^{-12}$ s for the doubly-excited configuration.   

Going further in analyzing the energy dependence of phase shifts and their interpretations in terms of time, split $\delta(E)$ into a slowly-varying $\delta_a(E)$ and more rapidly-varying $\delta_b(E)$, and use trigonometric identities for sines and tangents,

\begin{equation}
\sin^2 (\delta_a + \delta_b) = \sin ^2 \delta_a (-\cot \delta_a -\cot \delta_b)^2 /(1+ \cot ^2 \delta_b).
\label{eqn4}
\end{equation}
Defining a reduced energy $\epsilon = (E-E_r)/\frac{1}{2}\gamma = -\cot \delta_b$ and a background parameter $q = -\cot \delta_a$, the cross section in Eq.~(\ref{eqn4}) can be written in the form (Eq.(8.1) and Eq.(8.2) of \cite{ref1}, and Eqs.(5) and (6) of \cite{ref7})

\begin{equation}
\sigma = \sigma_0 + \sigma_a (q+\epsilon)^2 / (1+ \epsilon ^2), 
\label{eqn5}
\end{equation}
with $\sigma_0$ a background and one partial wave showing a resonance structure. 
The use of cotangents allows mapping of energy values from smaller than $E_r$ to larger into a variation of $\delta_b$ from 0 to $\pi$ as the cotangent goes from $-\infty$ to $\infty$. $\Gamma$ sets the scale for the rapidity in rise of the phase shift, the smaller the $\Gamma$ the more rapid the climb and narrower (in $E$) the rise and fall of $\sigma$, with an attendant longer auto-ionization lifetime. 

The above general form of the resonance cross section depends on three parameters, $E_r$, $\Gamma$, and $q$. These are the energy position, width, and ``Fano profile parameter" (may have either sign), respectively. The resonance structure is called a ``Beutler-Fano" profile and is asymmetric about $E_r$, with a zero at $\epsilon = -q$ \cite{ref8,ref9}. It reduces to a symmetric ``Breit-Wigner" form \cite{ref10} for large $|q|$, the resulting energy dependence then entirely in the Lorentzian $(1+\epsilon^2)^{-1}$. See, for instance, different profiles in Fig. 27 of \cite{ref11} and Fig. 1 in \cite{ref9}. In the interpretation of a quasi-bound state embedded in a background continuum, a quantum superposition results in constructive and destructive interference across the profile as one traverses the two wings of the resonance. The position of the zero, that is, completely destructive interference, occurs at an energy less than or greater than $E_r$ depending on the sign of $q$. Although not necessary in the above derivation, if $\delta_a$ is slowly-varying, $q$ is sensibly a constant independent of energy and even common to entire Rydberg series of resonances in a channel \cite{ref12}. This is often seen in atomic spectra. For this purpose, the decomposition of the identity in Eq.~(\ref{eqn4}) is instructive, the first factor of $\sin ^2 \delta_a$ smoothly varying and even possibly constant as the phase shift climbs through $\pi$ at successive values of $E_r$ along the series. But that factor and thereby $q$ itself may vary, perhaps even change sign as in ``$q$-reversal" \cite{ref13} along a series, depending on the variation of $-\cot \delta_a$ with energy. That could be seen as a further decomposition of that factor itself as per the same identity into smoother and more rapidly varying pieces.

Extending further into continuous variations, and as with lifetime defined from $\hbar$ with an energy interval $\Gamma$ in the denominator, time itself may be introduced into a stationary state analysis through a differential in energy. This is the Wigner-Eisenbud-Smith \cite{ref14,ref15} ``time delay" (section 4.4 of \cite{ref1})

\begin{equation}
\tau = 2 \hbar \,\, d\delta/dE, 
\label{eqn6}
\end{equation}
the factor of 2 reflecting the two-way traversal from infinity to small $r$ and back to infinity in scattering. The half-scattering interpretation would argue for dropping the factor of 2 when considering photoionization. The terminology of a time delay stems from interpreting the rise in $\delta$ through $\pi$ at a quasi-bound state as a temporary trapping into that configuration (such as $2s2p$) before the electron again emerges back in the $1sEp$ continuum. But, as with phase shift itself, its energy-derivative and, therefore, time delay may be of either sign. A drop in the phase shift with increasing energy would be a negative time delay or time ``advance." Interestingly, while time delay can take arbitrarily large values (the quasi-bound state almost bound), a causality argument limits how negative the advance can be \cite{ref14}. That magnitude cannot exceed twice the traversal time of the electron with speed $v$ of the range of the interaction. That limit is attained when the incident electron is repelled by an infinite hard-sphere repulsion at that point. With $e^2/\hbar = \alpha c$, where $\alpha$ is the fine-structure constant ($\approx 1/137$), setting a scale of atomic speeds, and $a_0 = \hbar^2/me^2$ an atomic unit of distance, this gives an atomic unit of time of $\hbar^3/me^4 \approx 45$ attosec.

The time delay $\tau(E)$ is itself energy-dependent, reflecting residual energy dependences beyond the first derivative in the phase shifts. Depending as it does on a derivative, it may be a more sensitive measure of energy dependence of scattering and its use has been advocated for this purpose not only at resonances but also in other circumstances. An example is in photoabsorption in the region of a ``Cooper minimum" in the cross section for many atoms and negative ions (section 4.5 of \cite{ref11} and section 12 of \cite{ref16}). Such a minimum in the photoionization cross section a little above threshold was first observed \cite{ref17} in alkali atoms and attributed to a change in sign of the dipole matrix element governing the process \cite{ref18,ref19}. It has been observed in many atomic systems and systematics of the phenomenon were established by Cooper \cite{ref20,ref21}. By convention, radial wave functions are defined as positive near the origin, and therefore the matrix element starts positive as well at threshold. At higher energies, however, the matrix element may be negative and thus a zero in between can cause an abnormally small cross section around that point. Among conditions necessary for this is that of the two dipole allowed $\ell \rightarrow \ell \pm 1$ transitions, the dominant one of $\ell +1$ exhibits this zero for absorption from $n\ell$ when the state $n, \ell+1$ is an allowed configuration but un-filled. This means that for photoionization from the $p$ shell of rare gases, Ne does not but higher rare gases show a Cooper minimum. Whereas $n=2$ does not have a $d$, larger $n$ of the higher rare gases do and it is unoccupied in the valence shell. Similarly, photoabsorption from the $3d$ of Kr does not but from $4d$ of Xe does display such a minimum with $4f$ allowed but unfilled. See Figs. 5 and 11 of \cite{ref16}. 

At energies just above threshold, photoabsorption differs radically for negative ions and neutral atoms. The Wigner threshold law \cite{ref22} applies for the former with $\sigma \propto E^{\ell +  \frac{1}{2}}$, thus starting at zero at the threshold for photodetachment. On the other hand, the photoionization cross section for neutral atoms is a constant at threshold, independent of $\ell$ because of the long range Coulomb field seen by the escaping photoelectron. A recent detailed study by Manson and collaborators \cite{ref23} through relativistic random phase calculations of Ar and its isoelectronic negative ion Cl$^- \, 3p \rightarrow Ed, Es$ documents these features, and especially dramatically the sensitivity of the time delay $\tau$. More so than the cross section $\sigma$, matrix element, or phase shift, it is $\tau$ that has the most pronounced structure in both Ar and Cl$^-$ around the Cooper minimum lying 45 eV above threshold. Near its photodetachment threshold, Cl$^-$ is very different with its $3p \rightarrow Es$ channel that dominates there starting at zero, leave alone the even more suppressed $3p \rightarrow Ed$. By contrast, in Ar, both begin with a finite value at threshold. But, in the vicinity of 30 - 50 eV, both Ar and Cl$^-$ are very similar in their Cooper minimum. The time delay is nearly constant (that is, phase shift flat in energy) around zero for several eV on either side but with a sharp drop, a time advance of several hundred attosec, in a narrow energy range. See Figs. 1-3 of \cite{ref23}. 

Turning from energy dependence, time elements can also be introduced into the dependence of phase shifts on angular momentum. First, the Wigner threshold laws referred to above may be viewed in terms of the angular momentum barrier experienced by the electron. In classical mechanics, for some energy $E$ or linear momentum $k$, the angular momentum involves in addition a lever arm so that higher $\ell$ values are kept further away in $r$ from the origin. This is more acute at small $E$ and $k$. Angular momentum appears correspondingly in the radial equation of quantum mechanics as an effective potential $\ell (\ell +1) \hbar^2/(2mr^2)$ that keeps the electron away from the origin. The higher the $\ell$, the larger this barrier and the more suppressed the wave function at small $r$ with an $r^{\ell}$ dependence. Especially for a low energy electron, the classical turning point $r_{\rm cl} = (\ell +\frac{1}{2}) \hbar /\sqrt{2mE}$ separates an inner and outer region and there is a similar suppression of the probability for the electron to tunnel past it from near the origin to larger $r$. Indeed, a JWKB expression for this tunneling probability gives precisely the $k^{2\ell +1}$ dependence of the phase shift that is the basis of Wigner threshold laws \cite{ref24}. 

The elastic scattering phase shift or $\tan \delta_{\ell}$ has such a dependence and its square enters in Eq.~(\ref{eqn3}) for $\sigma_{\rm el}$, elastic scattering involving two-way traversal from large to small $r$ and back. For inelastic cross sections, the $\ell$ of the outgoing partial wave similarly involves the $E^{\ell +  \frac{1}{2}}$ of their threshold behavior as noted above for photodetachment of Cl$^-$. However, in the presence of an attractive Coulomb field that is of longer range than $1/r^2$, and thereby is the dominant factor at low energies, the tunneling and dependence on $\ell$ drops out \cite{ref24}. This is also manifest in the zero-energy radial wave function's behavior, $J_{2\ell+1} (\sqrt{8Zr/a_0})$, that is energy-independent for all $\ell$. (For a repulsive Coulomb potential, the tunneling suppression is enhanced leading to an exponentially reduced cross section with the form $\exp(-c/\sqrt{E})$, with $c$ an energy-independent constant.)

In complex atoms beyond hydrogen, the angular momentum barrier takes on even richer structure upon combination with the Coulomb potentials felt by the electron. In hydrogen, that combined radial potential for the electron has a smooth, monotonic dependence on $r$ with a single minimum. In a more complex atom, the photoelectron in its passage from near the origin and nucleus of charge $Ze$ to asymptotic $r$ sees a complicated many-body interaction with all the other electrons that may be viewed as an effective potential that interpolates between the $-Ze^2/r$ at very small $r$ to $-e^2/r$ at infinity. In between, therefore, it falls faster and indeed can be well described by a faster $1/r^2$ that is also the dependence of the angular momentum contribution. The combination of nearly equal and opposite terms can, in some instances, lead to a near cancellation or even go positive for a certain range of $r$. This results in a ``two-valley" potential with an inner deep attractive well and a shallow outer well separated by an intervening positive barrier. This is, of course, more likely with larger $\ell$. That barrier can influence strongly low-energy electrons either incident from large $r$ in their penetration to the inner region or electrons on their passage out from the inner well to infinity. 

Already in the 1930s, Fermi \cite{ref25} and Mayer \cite{ref26} had linked a variety of atomic properties to this two-valley potential as, for instance, the nature of transition elements and lanthanides and actinides. The filling of the corresponding $d$ and $f$ shells is affected by the intervening barrier so that it is only when the barrier drops in La (57) as compared to Xe (54) that the $4f$ shell starts to fill and does so through the lanthanide series to Lu (71). All fourteen of those elements with the $f$ shell getting filled in the inner well are otherwise similar in the outer regions which is at the root of their occupying a single spot in The Periodic Table. In working with the Thomas-Fermi model as a one-electron potential across The Periodic Table, the barrier was observed for $f$ electrons but not $d$ \cite{ref26}. But, with the advent of numerical Hartree-Fock calculations across the elements, a Hartree-Slater (exchange) potential became available in the 1960s for all $Z$ and all $r$ \cite{ref27}. This is more accurate and displays atomic shell structure unlike the Thomas-Fermi potential. Even subtleties such as dips at half-filling of $d$ and $f$ shells are exhibited (see Fig. 1 of \cite{ref28}), again in conformity with observed properties along transition elements and lanthanides. The plot of an effective one-electron potential for all $Z$ as a function of $r$ showed these changes and were reflected in calculations of photoabsorption spectra displaying similar non-monotonic changes from one atom to the next and especially an extremum at Au \cite{ref29}.  The two-valley potential also showed non-monotonic dependences in $Z$ since the intermediate region is one of a balance between two nearly equal but opposite $1/r^2$ contributions, so that variations in either can be consequential (see Fig. 2 of \cite{ref28}). The two-valley potential occurs for both $d$ and $f$ electrons and has since been investigated extensively \cite{ref30}.

The barrier region at intermediate $r$ that separates the inner well, where the photoelectron is initially excited upon absorption from a ground or lower energy state, from the outer Coulomb well can at low energies suppress electron escape and thus the cross section. Generally, higher $\ell$ have larger cross sections both because of the higher statistical weight and because of larger overlap of radial wave functions. See, for instance, Fig. 19 of \cite{ref11}, Figs. 5 and 11 of \cite{ref16}, and \cite{ref31,ref32}. In the Ar example discussed above, the $3p \rightarrow Ed$ is suppressed just above threshold by the barrier, later to increase rapidly. The tunneling probability involves barrier width and height in an exponent so that changes in them with even small changes in $E$ can be reflected in large changes in the cross section that involves that exponential. In combination with the later fall off in $E$, the cross section shows a peak in such $3p \rightarrow Ed$, or in $4d \rightarrow Ef$ in Xe, and these are termed ``shape" resonances (section 4.5 of \cite{ref11}) to distinguish them from quasi-bound state or ``Feshbach" resonances discussed earlier \cite{ref33,ref34}. The latter occur just below a corresponding threshold ($E< 0$) whereas the former lie above ($E > 0$) and arise from the shape of the potential that gives them their name. The photodetachment of H$^-$ in the vicinity of the $n=2$ threshold at 10.95 eV provides a particulary nice example with one very sharp $^1P^{o}$ Feshbach resonance just below and a broad shape resonance just above \cite{ref35,ref36}.

With the advent of attosecond measurement technology through streak cameras \cite{ref37,ref38}, a recent experiment \cite{ref39} made a direct measurement of photoelectrons from tungsten diselenide WSe$_2$. Electrons from the conduction band and from the filled $3d$ shell of Se were recorded to arrive at different times at the detector. The $Ep$ electrons from that shell have no angular momentum barrier of a two-valley potential as do the $Ef$ photoelectrons. Interpreted as an extra tunneling time for them, values of $\approx$ 20 attosec were measured consistent with that expected for tunneling through the intervening barrier \cite{ref39}. As in Fig. 2 of \cite{ref28}, the $d$-wave barrier builds up from Sc to the completion of the $3d$ shell at Cu and remains for the subsequent elements through Se to Kr.

Alternatives to the streak camera can also reach attosec scales such as direct measurement of the phase of the photoionization amplitude with free electron lasers \cite{ref40}. All these measurements now possible of very short time scales allow the probing of Fano resonances in the time domain, following their very build up \cite{ref41,ref42,ref43,ref44} in time. See also a pedagogical treatment in \cite{ref45}. Remarkably, even shorter time scales can now be measured. While the time for an atomic electron to travel the length of a Bohr radius is, approximately, 45 attosec, the corresponding light travel time is shorter by the fine-structure constant, that is, of the order of 300 zeptosec. This transit time between the two hydrogen atoms in the H$_2$ molecule has recently been measured \cite{ref46}.  

\section{Hybrid descriptions and the role of technology}

Our description of energy and time as two alternative representations may be seen as limiting extremes, and one can envisage hybrids in between. Indeed, in sound and acoustics, besides the Fourier conjugates of frequency and time domains, western music notation had invoked from its beginnings a representation/language that involves both. Notes of frequencies within an octave on some tempered scale are also grouped in a linear time sequence on the page. Correspondingly, finite wave trains or ``wavelets" have also a long history in mathematical physics \cite{ref47,ref48,ref49}. In atomic structure calculations too, uses of such basis sets or of B-splines, even if over-complete, have been shown to improve convergence \cite{ref50}. 

An aspect of which of energy or time representations to use is the role played by what is technologically convenient or indeed possible. This is part of a more general theme of the importance of technology in making advances in fundamental science, sometimes forgotten against the more commonly advanced claims made for basic science leading to new technologies. That is indeed so, but equally, available technologies of the day drive what can even be investigated and how. Thus, with the estimate above of a few ten attosec for photoelectrons to tunnel through angular barriers or escape the atom, direct measurements of such short times were outside the reach of experiment in the first decades of the field. In its place, more accurate energy measurements and higher resolution and time delay as per Eq.~(\ref{eqn6}) allowed such access. The advent of very short laser pulses, first on the femto second scale of molecular transformation and then of attoseconds, along with streak cameras and such techniques, have only recently permitted such direct access to the very short times involved as noted above. Angular momentum barriers that were first mapped out \cite{ref28} in the mid-1960s had to wait a half-century for direct measurements of tunneling time through them \cite{ref39}. Already, even that mapping became possible only because of early computer calculations \cite{ref27} such as Hartree-Fock-Slater that made possible study across the Periodic Table with more accuracy than the Thomas-Fermi model of the previous 30 years. And for theoretical calculations, the very high speed numerical calculations now feasible with supercomputers allow meaningful direct integration of the time-dependent Schr\"{o}dinger equation with very small time steps, without being restricted to time-independent calculations with basis state expansions for the stationary states of fixed $E$.                                        
       
The alternative descriptions of phenomena as evolution in time or as a single snapshot at an instant exist also outside of physics in other sciences and even general intellectual activity and philosophy. Astronomy, in particular, always had to resort to the latter because of the great mis-match between time scales for evolution of stars, galaxies and the entire Universe as compared to the human lifespan. Instead of following one star or galaxy from birth to death, one observes an entire collection as available at this particular epoch, capturing various similar objects at different stages of their evolution. The finite, and different, times it takes for light to reach us helps in this regard. In biology, today we can through fast cameras and time-lapse photography follow a single moth or butterfly from egg through larva to caterpillar, pupa, and emergent insect. (In some butterflies such as the monarch, there are even several cycles of the flying insect stage between the first and the return to the breeding trees in Mexico from the in-between wanderings over North America.) But, as an alternative to following an individual, botanical and zoological illustrations can capture on one plate the entire life cycle as depicted simultaneously by different individuals. An early example were the canvases of the N\"{u}rnberg artist, Maria Sybille Merian, who uniquely for her time painted such life cycles embracing caterpillar and moth that she had seen in Surinam \cite{ref51,ref52}.  
  
\section{Acknowledgments}
The perspectives one develops of a field owe much to several discussions with colleagues over many years. My studies in photoabsorption began in the first days of graduate school under the tutelage of Prof. Ugo Fano. Just weeks later, I met and started discussing with Steve Manson, an interaction that has continued for over fifty years. This article in honor of his contributions to the field is also offered as a tribute to others with whom I discussed from those first months and extending over decades but who have now, sadly, passed away: Philip Altick, John Cooper, Ugo Fano, Mitio Inokuti, Yong-Ki Kim, and Anthony Starace.

\end{document}